\documentclass[aps,pra,onecolumn,floatfix,notitlepage,superscriptaddress]{revtex4-1}
\usepackage{bm}
\usepackage{graphicx}
\usepackage{amsmath}
\usepackage{amssymb}

\usepackage{epstopdf}
\usepackage{subfigure}

\usepackage{color}

\begin{document}
\title{Surface and electronic structure of SmB$_6$ through
Scanning Tunneling Microscopy}

\author{S. R\"{o}{\ss}ler}
\affiliation{Max Planck Institute for Chemical Physics of Solids,
01187 Dresden, Germany}
\author{Lin Jiao}
\affiliation{Max Planck Institute for Chemical Physics of Solids,
01187 Dresden, Germany}
\author{D.~J. Kim}
\affiliation{Department of Physics and Astronomy, University
of California, Irvine, CA 92697}
\author{S. Seiro}
\affiliation{Max Planck Institute for Chemical Physics of Solids,
01187 Dresden, Germany}
\thanks{Present address: Department of Chemistry and Physics of 
Materials, University of Salzburg, Hellbrunner Str. 34, 5020 
Salzburg, Austria}
\author{K. Rasim}
\affiliation{Max Planck Institute for Chemical Physics of Solids,
01187 Dresden, Germany}
\author{F. Steglich}
\affiliation{Max Planck Institute for Chemical Physics of Solids,
01187 Dresden, Germany}
\author{L.~H. Tjeng}
\affiliation{Max Planck Institute for Chemical Physics of Solids,
01187 Dresden, Germany}
\author{Z. Fisk}
\affiliation{Department of Physics and Astronomy, University
of California, Irvine, CA 92697}
\author{S. Wirth}
\email{Email: wirth@cpfs.mpg.de}
\affiliation{Max Planck Institute for Chemical Physics of Solids,
01187 Dresden, Germany}
\date{\today} 

\begin{abstract}
SmB$_6$, a so called Kondo insulator, is recently discussed as
a candidate material for a strong topological insulator. We
present detailed atomically resolved topographic information
on the (001) surface from more than a dozen SmB$_6$ samples.
Atomically flat, {\it in situ} cleaved surfaces often exhibit
B- and Sm-terminated surfaces as well as reconstructed and
non-reconstructed areas {\it coexisting} on different length
scales. The terminations are unambiguously identified. In
addition, electronic inhomogeneities are observed which likely
result from the polar nature of the (001) surface and may
indicate an inhomogeneous Sm valence at the surface of SmB$_6$.
In addition, atomically resolved topographies on a (110) surface
are discussed.
\end{abstract}

\maketitle

\section{Introduction}

Topologically non-trivial surface states have evolved into a
highly topical and active field of research. Here, the interest
stems not only from the physics involved \cite{rmp10} but also
from some of the unusual transport properties of the electrons
within the surface states which render them attractive for
potential applications \cite{Ste15}. The main idea relies on
a sufficiently strong spin-orbit coupling in a bulk insulator
which can result in topologically protected (by time-reversal
symmetry) surface states \cite{Kan05,Fu07}. To this end, the
product of the parity of all involved bands at the time
reversal invariant momentum points is required to be $-1$. In
a seminal paper it was demonstrated that such conditions could
be fulfilled in strongly correlated 4$f$ systems in which a
hybridization between conducting $d$-bands and localized
$f$-bands takes place \cite{Dze10}. Here, the so-called Kondo
insulators \cite{Aep92,Ris00} are of particular interest since
they are characterized by a narrow gap and strong spin-orbit
coupling. Subsequent band structure calculations for the
well-known Kondo insulator SmB$_6$ indeed predicted
non-trivial surface states \cite{Tak11,Lu13,Ale13,Kim14}. It
also sparked a flurry of experiments. Transport measurements
were able to show that there is surface conductance
\cite{Wol13,Zha13,Kim13}. A number of angle-resolved
photoemission spectroscopy (ARPES) studies were conducted in an
effort to provide support to the claim that these surface states
are in fact topological in nature. In particular, results of
spin-resolved ARPES were interpreted as evidence for the latter
\cite{Xu14,Sug14}. However, other views have also been put
forward including the involvement of a surface reconstruction
and of the polar nature of the (001) SmB$_6$ surface
\cite{Zhu13,Fra13}. In addition, ARPES results down to 1 K have
been discussed in terms of Rashba splitting \cite{Hla15}.

One severe obstacle in resolving the afore-mentioned issues
is the complexity of the SmB$_6$ surfaces, even if they are
cleaved {\it in situ} and smooth down to an atomic level.
We apply Scanning Tunneling Microscopy (STM) and Spectroscopy
(STS) to provide information on the atomic length scale to
show the multitude of different terminations often coexisting
on the same surface. Atomically resolved spectroscopic data
indicate electronic inhomogeneities on non-reconstructed
surfaces. In addition, signatures of the Kondo effect being
at play are presented.

\section{Experimental}

Single crystals of SmB$_6$ were synthesized by using an Al
flux method \cite{Kim13}. The STM/STS work was conducted
mostly in a low-temperature (LT) STM (base temperature 4.6 K,
Omicron Nanotechnology) with ultra-high vacuum (UHV) of $p \leq
3 \! \times\! 10^{-9}$ Pa. Also, a UHV cryogenic STM was
employed providing a base temperature of 0.3 K and a magnetic
field of up to 12 T. The SmB$_6$ single crystals were cleaved
{\it in situ} of the respective STM chamber and at temperatures
of about 20 K. We report results on a total of 18 cleaves.
Tunneling spectroscopy was measured by using a lock-in
technique with small bias modulation ($V_{\rm mod}$) at
117 Hz. If not stated otherwise, topographic images shown
in the following were obtained at temperatures $T \sim 5$ K.

\section{Surface topography, $\{$001$\}$ crystallographic planes}
\label{100}
\subsection{Overview}
\label{sec-over}
\begin{figure}[t]
\begin{center}
\includegraphics[width=13.8cm]{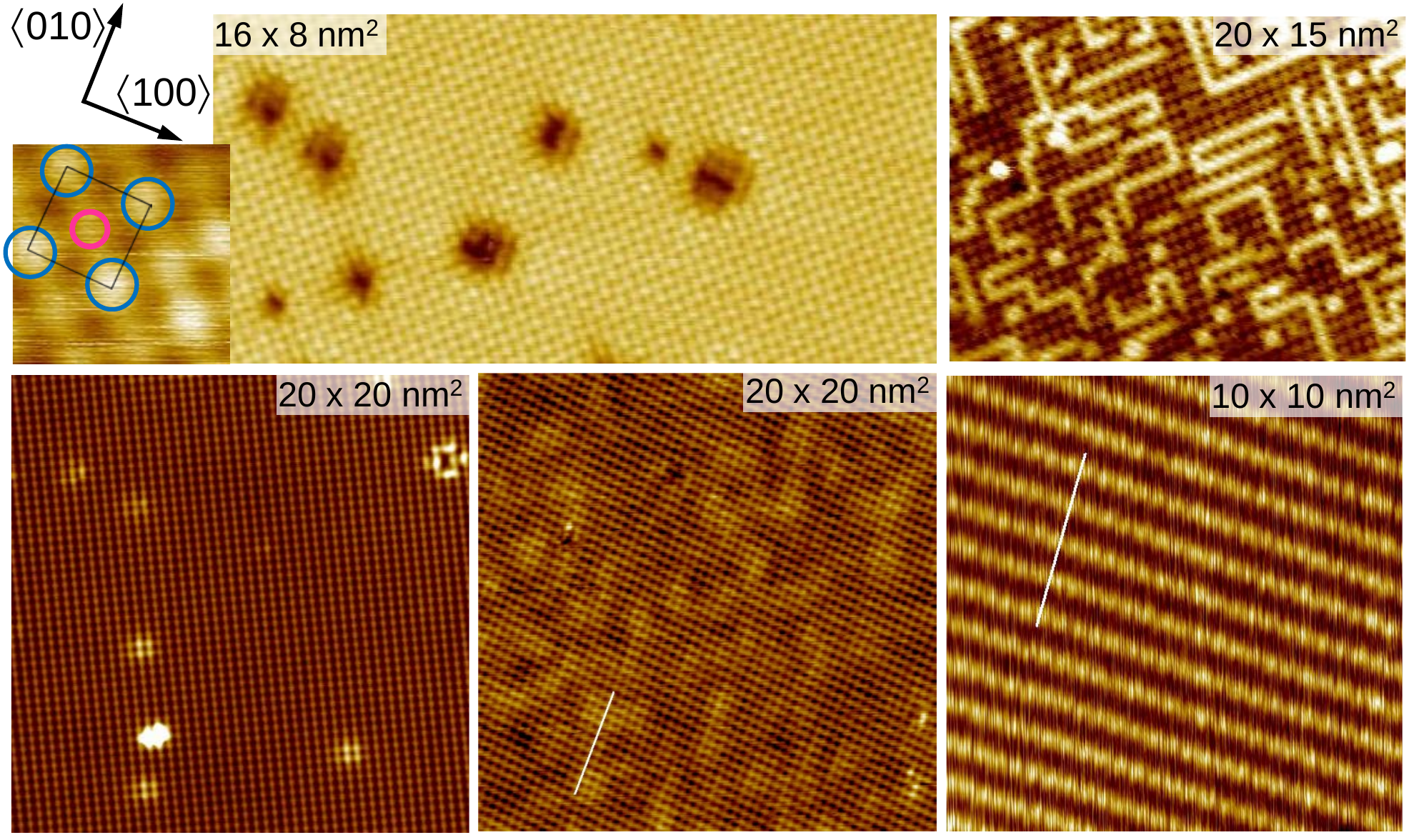}
\caption{Overview of the different topographies encountered
for $\{001\}$ planes. The upper left and center topographies
present Sm-terminated surfaces; the main crystallographic
directions are indicated. The zoom on the left shows an area
of $1 \times 1$ nm$^2$, with the Sm (blue) and B (pink) apex
atoms marked by circles \cite{Roe14}; cf. figure \ref{term}
for the crystal structure. The bottom left and center
topographies exhibit B-termination; the center one
adumbrates electronic inhomogeneity. The two images on the
right depict reconstructed surfaces; the upper one disordered,
the lower one ordered ($2\!\times\! 1$ reconstruction). White
lines indicate a main crystallographic direction.}
\label{topos} \end{center}
\end{figure}

In figure \ref{topos} an overview of the different surfaces
encountered so far is presented. It should be emphasized that
these different topographies may come in various sizes (they
can be as small as a few nm) and may coexist on one and the
same surface depending on the investigated location. In STS,
spectroscopy can be conducted on a well defined topography
while measurements in which a certain area is investigated (e.g.
ARPES, optical conductivity or point contact measurements) may
very well average over different terminations with unknown weight.
The upper central image of figure \ref{topos} presents a
Sm-terminated surface. This can most easily be recognized by
considering the main
crystallographic directions (shown on the left side): the lines
of corrugations run along the diagonal. This is obvious from the
zoom on the left \cite{Roe14} which shows an area of $1\!\times
\! 1$ nm$^2$. The Sm atoms (blue circles) exhibit the expected
square arrangement along the main crystallographic directions
and with correct distances (SmB$_6$: cubic crystal structure
with lattice constant $a =$ 4.133 {\AA}, cf. figure \ref{term}).
In addition, also the apex of the B octahedra (pink) is clearly
seen in the topography (the assignment of the atoms will become
clear below). Very similar surface structures were observed on
SmB$_6$ \cite{Rua14} as well as LaB$_6$ \cite{Ozc92}. In
contrast, the two topographies at bottom left and center depict
B-terminated areas. Here, the corrugations are aligned parallel
to the $\langle$100$\rangle$ directions (indicated by a white
line in the center image). However, the center image clearly
shows some underlying inhomogeneities (also seen in
\cite{Yee13}) which are \emph{not} present in the left image.
We consider this inhomogeneity to be electronic in origin and
discussed this below. The two images on the right visualize
differently reconstructed surfaces. The upper one shows that
the Sm atoms on top of the B surface do not necessarily need to
order into a ($2\! \times\! 1$) reconstruction as it is
presented in the lower image (for further discussion, see section
\ref{sec-reco}). Some of the surface terminations reported here
have also been observed by others \cite{Yee13,Rua14}. It
should be noted, however, that all these smooth terminations
need often to be searched for, while many areas appear rough
on an atomic scale.

\subsection{Assignment of the surface termination}
\begin{figure}
\begin{center}
\includegraphics[width=13.8cm]{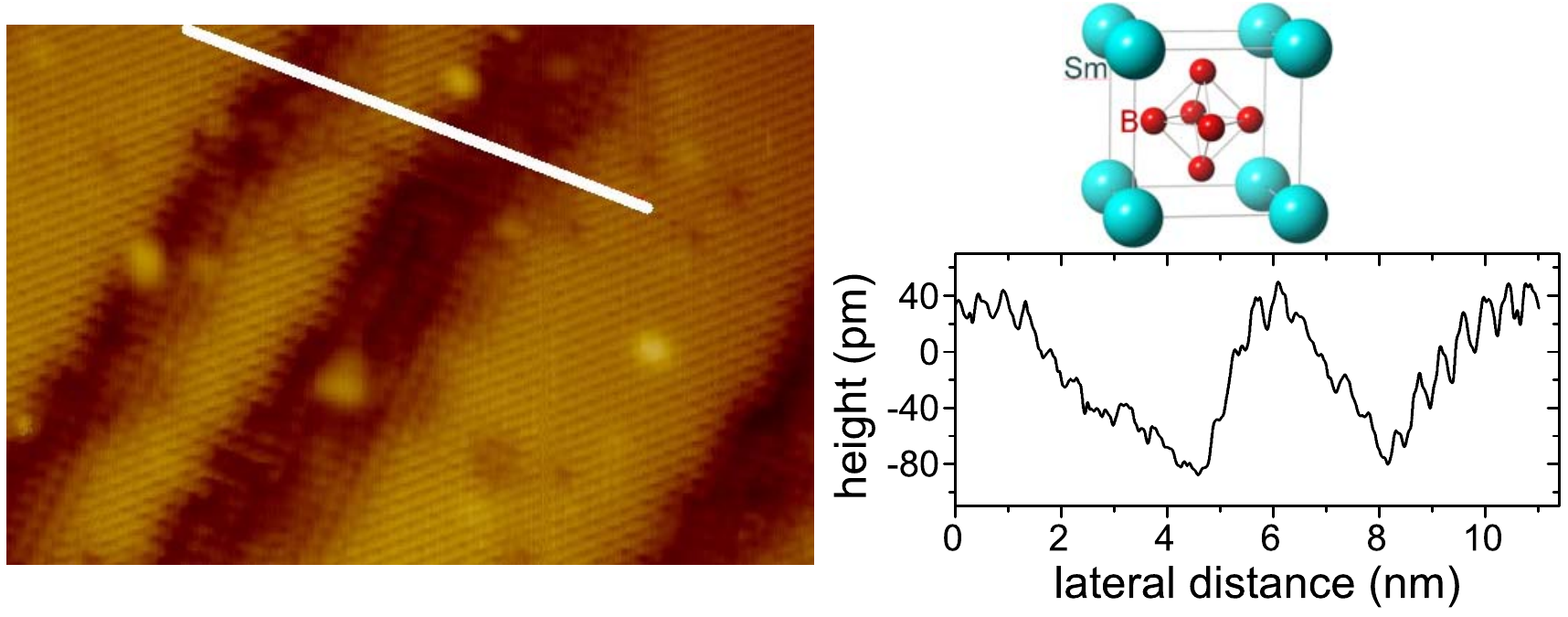}
\caption{Left: Example of non-reconstructed surface ($12
\times 18$ nm$^2$) showing several steps between Sm- and
B-terminated areas; bias voltage $V_b = 0.2$ V, set point
current $I_{sp} =$ 0.6 nA. The white line is almost parallel
to a $\langle$100$\rangle$ crystallographic direction. Right:
Height scan along the white line marked in the topography
image. The cubic crystal structure of SmB$_6$ is also shown.}
\label{term} \end{center} \end{figure}

In all discussions about the different surfaces it is
pivotal to assign the termination properly, i.e. to
unambiguously distinguish Sm- from B-termination. The likely
most reliable way to achieve this is by the observation of
atomically resolved steps separating different terminations,
and specifically if these step heights are in agreement with
expectations from the crystal structure \cite{Roe14}. Here,
further support to this assignment is presented; this time
by considering steps between alternating surface terminations
as presented in the topography figure \ref{term}, left. In
this $12 \times 18$ nm$^2$ field of view, three ``trenches''
of approximately 100 pm in depth can clearly be made out, see
height scan (figure \ref{term}, right) along the white line
marked in the topographic image. Note that this height scan
was taken almost parallel to one of the main crystallographic
directions, i.e. along $\langle$100$\rangle$. We assign the
upper (brighter in figure \ref{term}, left) terraces to
Sm-terminated areas and the lower (darker) ones to B
terminations for the following two reasons: i) The height
difference of the upper and lower terraces corresponds
approximately to the distance between the center of a B
atom and the plane spanned by the four closest Sm atoms,
i.e. half the distance of two \emph{inter}-octahedral
borons (0.083 nm). Of course, the heights as measured by
STM are also influenced by the local density of states (DOS)
and possibly also by some relaxation of the surface atom
distances with respect to their bulk distances. Nonetheless,
the measured heights are difficult to explain for any
other assumed surface configuration. ii) As already seen
for the non-reconstructed surfaces in figure \ref{topos},
the corrugations on the upper terraces run along lines
parallel to $\langle$110$\rangle$ (note that the white
line in figure \ref{term}, left is almost parallel to
$\langle$100$\rangle$). These corrugations have slightly
alternating heights corresponding to an alternating
visualization of Sm and B atoms with distances of
$\approx \frac{1}{2} \sqrt{2}a$ along the diagonal
\cite{Roe14}. In contrast, the corrugations within the
lower areas of figure \ref{term} are oriented along
$\langle$100$\rangle$ with distances of $a$ as expected
if apex atoms of the B octahedra are visualized.

Within the cubic crystal structure of SmB$_6$, the
octahedra form sturdy B$_6^{2-}$ polymeric anions. Yet, the
\emph{inter}-octahedral B-distance (0.1669 nm) is slightly
smaller than the \emph{intra}-octahedral distance (0.1744
nm) \cite{Mac13}. Hence, one might consider cleaving through
the B$_6$ octahedra rather than between them. Indeed, we
rarely observed ``doughnut-shaped'' structures (see figure
\ref{reco}(h)), similar to those reported in \cite{Rua14}.
These ``doughnuts'' have diameters of about one lattice
constant and therefore, are likely made up by eight boron
atoms, i.e. two respective atoms out of each of four
adjacent octahedra, cf. figure \ref{term} for the SmB$_6$
unit cell and \cite{Rua14}. In order to rationalize the
experimental findings, i.e. compare cleaving by breaking
inter-octahedral vs. intra-octahedral bonds, electronic
structure calculations with respect to the surface energies
$\gamma$ of differently terminated slabs were carried out.
Density functional theory calculations in the generalized
gradient approximation (GGA) were conducted for LaB$_6$ and
CaB$_6$, i.e. two hexaborides with integer valence La$^{3+}$
and Ca$^{2+}$ next to the intermediate valence $\nu \sim$
2.6 of SmB$_6$. The implementation in the all-electron first
principles code FHI-AIMS was employed using localized,
numerically tabulated atom-centered orbitals \cite{Blu09}.
For LaB$_6$ the calculated values are in good agreement with
\cite{Uij06}: an energetic preference of the surface formed
by cutting the inter-octahedral boron bonds ($\gamma = 2.97$
Jm$^{-2}$), as compared to surfaces that cut intra-octahedral
bonds ($\gamma = 3.62$ Jm$^{-2}$), was found. For the case
of CaB$_6$, the difference is somewhat larger (2.77 Jm$^{-2}$
and 3.92 Jm$^{-2}$, respectively). Surface relaxations do not
yield substantial changes and decrease all mentioned surface
energies by 0.1--0.2 Jm$^{-2}$. Assuming that qualitatively
the same picture holds for SmB$_6$, the calculations support
the assertion above that in rare cases, e.g. in concert with
respective lattice imperfections, the cleavage can also take
place between intra-octahedral bonds, while in the
\emph{typical} case the cleaving is expected to leave the B
octahedra intact.

The surface structure of figure \ref{term} can be
compared to the ($2\!\times\! 1$) reconstructed surface
(figure \ref{topos} lower right): In both cases, Sm and
B terminated surfaces are exposed simultaneously but on
different length scales. Yet, tunneling spectra obtained
on reconstructed surfaces clearly differ from those on
non-reconstructed ones \cite{Yee13,Roe14}. This again
emphasizes the need for detailed \emph{local} information
on the exact surface structure.

\subsection{Atomically reconstructed surfaces}
\label{sec-reco}
As already mentioned, a bulk truncated $\{$001$\}$ surface
of SmB$_{6}$ is polar due to uncompensated charges at the
\begin{figure}[t]
\centering\includegraphics[width=13.8cm,clip]{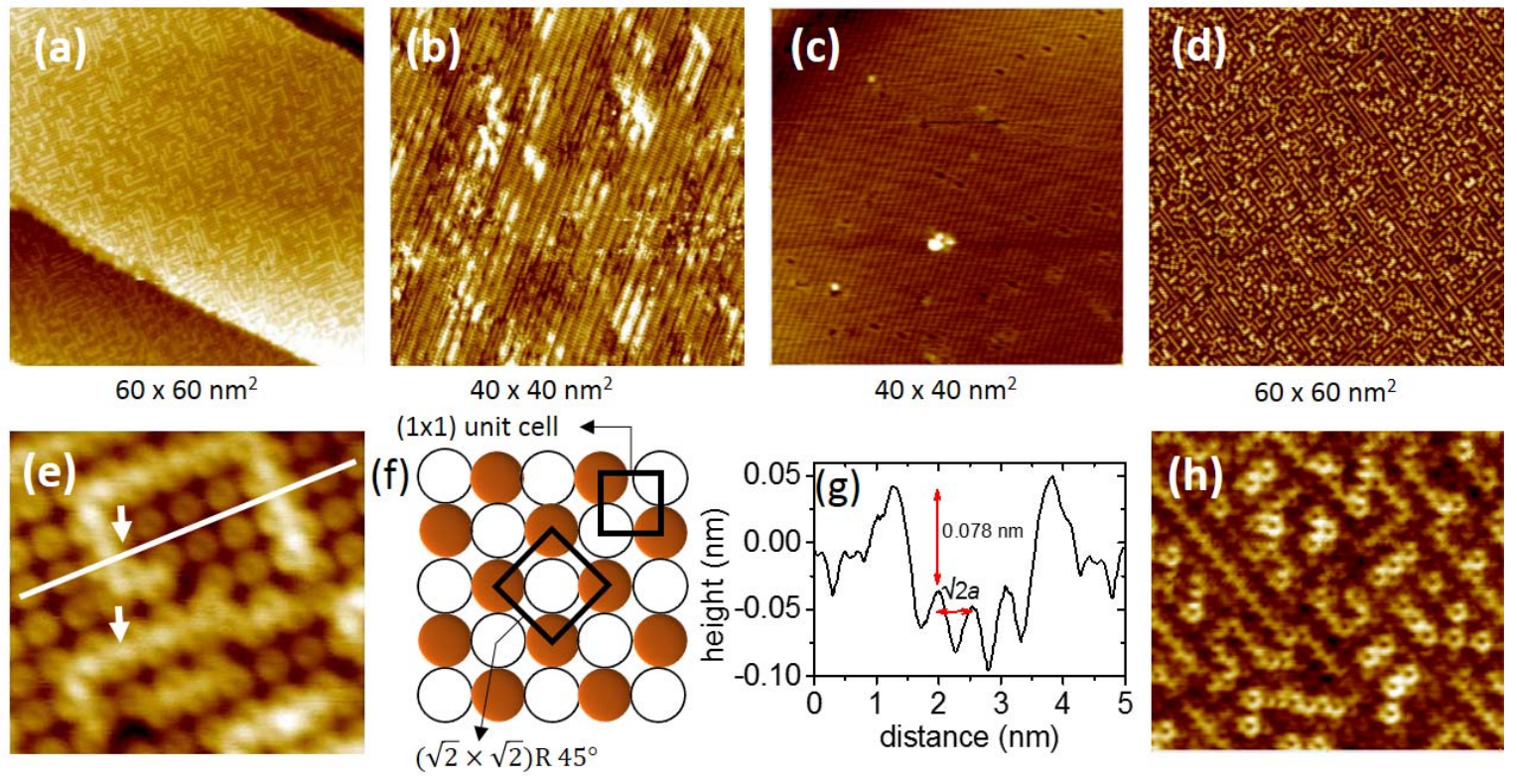}
\caption{(a)-(d) STM images of cleaved $\{$001$\}$ surface
of SmB$_{6}$. While (a), (b), and (d) are disordered, an
ordered (2$\times$1) reconstruction can be recognized in (c).
(e) High resolution image ($5\!\times\! 5$ nm$^2$) of a
disordered surface similar to (a), where Sm atoms form a
self-organized labyrinth on top of a
$(\sqrt{2}\times\sqrt{2})\mathrm{R}45^{\circ}$ reconstructed
B-terminated surface. Notice that across the Sm chains
B$_{6}$ octahedra are anti-phase with each other (for
example, see white arrows). (f) Schematic diagram of a $(\sqrt{2}\times\sqrt{2})\mathrm{R}45^{\circ}$ reconstructed
B-terminated surface. Empty black (solid brown) circles
represent missing (filled) B$_{6}$ octahedra. (g) Line scan
along the white line in (e). (h) High resolution zoomed
($10\!\times\! 10$ nm$^2$) image of (d) displaying
doughnut-like features.} \label{reco}
\end{figure}
surface. The intermediate valence of Sm renders a simple
electron counting even more complicated. However, several
reasons may give rise to complex surface morphologies, such as
a concomitant reduction of the free energy \cite{Rod93} and/or
an accommodation of the strain induced by the atomic size
mismatch of the constituents \cite{Tre95}. In consequence,
a variety of differently reconstructed surfaces was reported 
based on atomically resolved STM \cite{Yee13,Roe14,Rua14}. 
In figure \ref{reco}(a)-(d), the STM images of $\{$001$\}$ 
surfaces on areas as large as $60 \times 60$ nm$^2$ are 
presented. Such atomically reconstructed regions can extend up 
to a few microns. A single cleave at low temperatures can 
produce all these different types of topographies indicating 
that they all lie close in energy. The topographic images in
figures \ref{reco}(a), (b) and (d) are disordered, whereas
(c) displays an ordered (2$\,\times \,$1) reconstruction
(cf. also figure \ref{topos} right). The latter type of
reconstruction was observed previously in STM and low-energy
electron diffraction (LEED) measurements \cite{Aon78,Miy12}.

The surface displayed in figure \ref{reco}(a) is
particularly interesting: Here, Sm atoms form a
self-organized labyrinth-like structure on a
$(\sqrt{2}\times\sqrt{2})\mathrm{R}45^{\circ}$ reconstructed
B-terminated surface. In figure \ref{reco}(e), a high
resolution image of a similar surface is presented. Across
the Sm-chain, the B$_{6}$ octahedra are shifted by half a
unit cell, i.e., they are anti-phase with respect to each
other. In figure \ref{reco}(f) a schematic diagram of such a $(\sqrt{2}\times\sqrt{2})\mathrm{R}45^{\circ}$ reconstruction
is presented, in which the missing B$_{6}$ octahedra are
represented by black circles. These reconstructions are also
found in the mixed-valent material Fe$_3$O$_4$ \cite{Sta20}.
A height scan along the white line in figure \ref{reco}(e)
can be seen in figure \ref{reco}(g). The Sm--Sm valley-height
with respect to that of B$_{6}$--B$_{6}$ is found to be 0.078
nm, which is comparable to the distance of 0.083 nm between
the top B-atom of the B$_{6}$ octahedra and the plane made
up by the four closest Sm atoms in the SmB$_{6}$ unit cell
(i.e., half the inter-octahedra distance). Further, the
B$_{6}$-B$_{6}$ distance 0.56 nm is close to $\sqrt{2}a$,
which is consistent with a
$(\sqrt{2}\times\sqrt{2})\mathrm{R}45^{\circ}$-type
reconstruction. Figure \ref{reco}(h) depicts a high
resolution zoom-in image of (d). This disordered surface is
likely comprised of both Sm as well as B atoms. Similar
doughnut-like features have been reported in \cite{Rua14}.

It is needless to mention that the surface reconstructions
may easily result in even more complex surface states, which
make a straightforward observation of the topologically
non-trivial surface states an experimental challenge. A
well-known example for the formation of a metallic surface
layer is the Si(111)7$\times$7 reconstruction \cite{Yoo02}.

\section{Surfaces along the $\{$110$\}$ plane}
\label{110}
In contrast to the $\{$100$\}$ surface discussed in section
\ref{100}, the $\{$110$\}$ surface of SmB$_6$ is \emph{not}
polar. Hence, one could expect that an interpretation of
\begin{figure}[t]
\centering\includegraphics[width=9.8cm,clip]{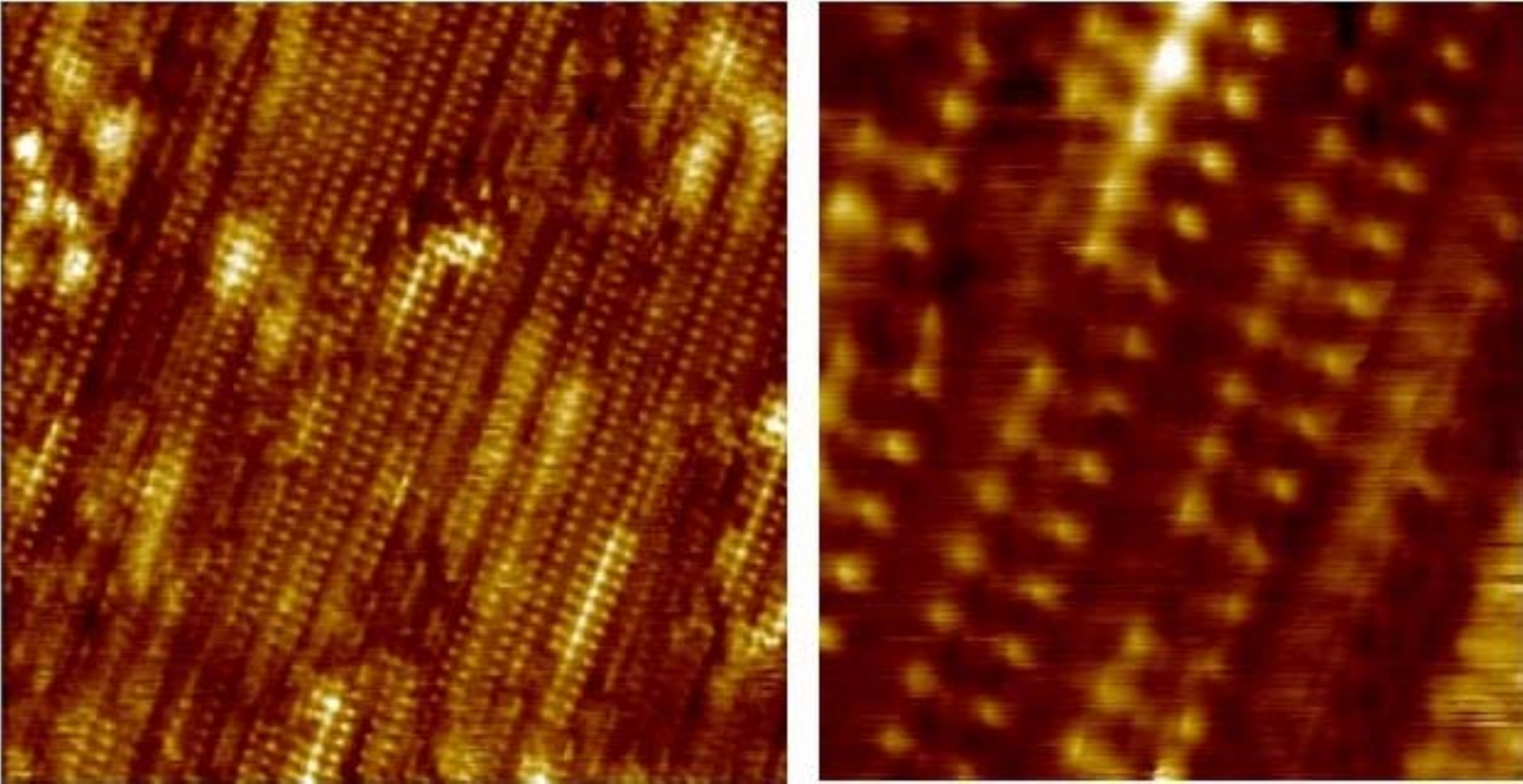}
\caption{STM images of a cleaved $\{$110$\}$ surface
of SmB$_{6}$; left: $20\!\times\! 20$ nm$^2$, right:
$4\!\times\! 5$ nm$^2$; $V_b = 0.2$ V, $I_{sp} =$ 0.6
nA. Although atomic corrugations can clearly be
recognized, these surfaces are not as smooth as those
obtain for $\{$001$\}$ planes.} \label{topo110}
\end{figure}
results obtained on the latter is more straightforward.
For the $\{$110$\}$ surface, two additional surface Dirac
points were predicted \cite{Ye13}. Also, experimental
results obtained on cleaved $\{$110$\}$ surfaces have been
reported \cite{Che15,Tan15}.

Different areas of a cleaved $\{$110$\}$ surface are
presented in figure \ref{topo110}. Clearly, atomic
resolution is achieved and the rectangular arrangement
expected for Sm atoms on a $\{$110$\}$ plane with correct
distances is seen in the zoomed-in image. However, the
surfaces appear certainly not as smooth as those obtained
on $\{$001$\}$ surfaces, contain numerous defects and
therefore, likely, different types of atoms and may
exhibit locally varying atomic environments.
Consequently, care has to be taken in interpreting
spectroscopic results unless the quality of the
surfaces is improved.

\section{Electronic inhomogeneities at the surface}
\label{inhomo}
In the following we focus on $\{$001$\}$ surfaces which are
not atomically reconstructed. Spectroscopy on smaller (a few
\begin{figure}[t]
\centering \includegraphics[width=15.6cm,clip]{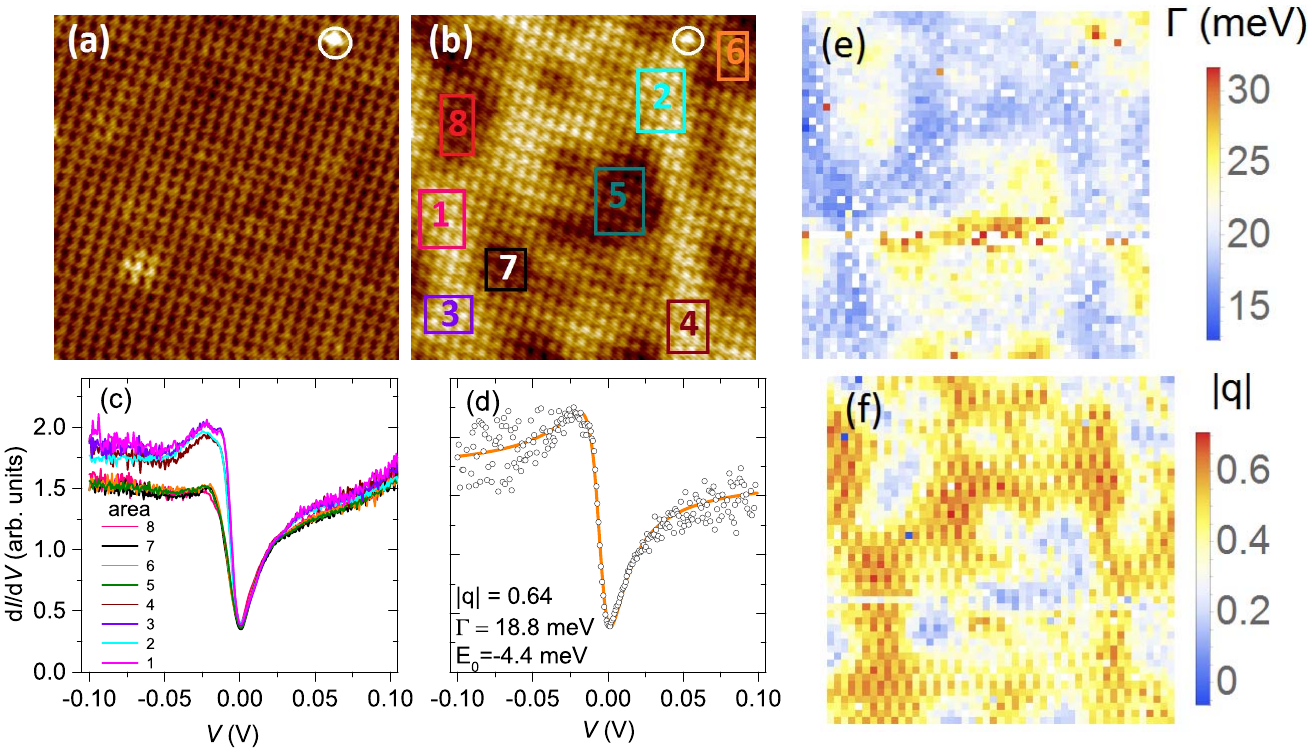}
\caption{Topography of SmB$_6$ over an area of $10\!
\times \! 10$ nm$^2$ measured at 6~K using a dual bias
mode: (a) Forward scan with $V_{b} = +0.025$~V and (b)
backward scan with $V_b = -0.025$~V. An impurity marked
by white circles confirms that both images were taken
on the same area. (c) Average tunneling conductance $g(V)$
on areas marked by matching colors in (b). (d) Typical
single tunneling spectrum and corresponding fit (continuous
line) to Equation \ref{Fano}. (e) and (f) Spatial variation
of $\Gamma$ and $|q|$, respectively.} \label{inhom}
\end{figure}
nm in extent) non-reconstructed areas revealed indications
for the hybridization gap \cite{Roe14} expected for a
Kondo insulator at low temperatures. In contrast, on
larger areas---at least several tens of nm in extent---the
tunneling spectral line-shape typically appeared as a
Fano resonance \cite{Fano61}, owing to a quantum mechanical
interference of the electrons tunneling from (to) the STM
tip into (out of) the quasiparticle states and the
conduction band. Then, the tunneling conductance
$g(V)$ = d$I(V)$/d$V$ can be fitted by \cite{Sch00}
\begin{equation}
g(V)\propto\frac{(\epsilon + q)^2}{\epsilon^2+1}\; , \qquad
\epsilon = \frac{2(eV-E_0)}{\Gamma}\; . \label{Fano}
\end{equation}
Here, $\Gamma$ and $E_0$ are the width of the resonance
and its position in energy, respectively. The asymmetry
parameter $q$ is determined by the probabilities of
tunneling into the quasiparticle states and the conduction
band, as well as by the particle-hole asymmetry \cite{Fig10}.
Although the bare local DOS of SmB$_{6}$ is obscured by the
Fano resonance, the parameter $\Gamma$ can be related to
the Kondo temperature $T_{K}$ \cite{Naga02}.

The STM topography, as shown in figure \ref{inhom}(a), was
conducted on an area of $10 \times 10$ nm$^2$ in a dual
bias mode. This implies that the same area can be scanned
with two different bias voltages applied for the forward
and backward scans. In the present case, the forward scan
was conducted with $V_b = +0.025$~V, figure \ref{inhom}(a),
while $V_b = -0.025$~V was used for the backward scan in
figure \ref{inhom}(b). The fact that exactly the same
areas were visualized is evidenced by the impurity marked
by the white circles in figures \ref{inhom}(a) and (b).
Nonetheless, a strong spatial inhomogeneity is seen in
figure \ref{inhom}(b), which is not present in figure
\ref{inhom}(a). Since the tunneling conductance is a
convolution of tip--sample distance and the local DOS,
this suggests not only a spatially inhomogeneous but
also a asymmetric local DOS with respect to the sign of
$V_b$. This can be more clearly inferred from the
tunneling conductance presented in figures \ref{inhom}(c).
Here, the d$I$/d$V$-curves were obtained by averaging
over the areas shown in (b) with matching colors. These
curves differ significantly between $-0.2 \lesssim V_b <$
0 mV. A fit of all 2500 individual spectra obtained within
the field of view of figures \ref{inhom}(a) and (b) were
fitted to Equation \ref{Fano}; an example is given in
figure \ref{inhom}(d). The resulting spatial maps for
$\Gamma$ and $|q|$ are presented in figures \ref{inhom}(e)
and (f), respectively. Notably, $\Gamma$ and $|q|$ are
anti-correlated, i.e., the regions with larger values of
$\Gamma$ exhibit smaller values of $|q|$. This is consistent
since larger values of $\Gamma$ suggest stronger
hybridization. Because the hybridization in SmB$_{6}$ takes
place between Sm 4$f$ states and the 5$d$ conduction band,
a large $\Gamma$ may implies that the 5$d$ bands are more
occupied, which in turn can give rise to a higher
probability of tunneling into the conduction band, i.e.
smaller $|q|$. We speculate that the spatial inhomogeneities
of $\Gamma$ and $|q|$ is related to a spatially inhomogeneous
intermediate valence of Sm at the surface of SmB$_{6}$.

Within our field of view, $E_{0}$ is found to vary between
$-7$ meV $<E_0 < $ 1~meV, with a pronounced maximum of
probability at $E_{0}\approx -3.8$~meV. Since the
hybridization gap in SmB$_{6}$ is about 15--20~meV
\cite{Zha13,Roe14,Gor99,Neu13}, one may assume that the
Fano resonance found here is likely a phenomenon related
to the in-gap states lying close to the Fermi level
$E_{\rm F}$. As mentioned above, in order for the Fano
resonance to show up, two types of tunneling channels are
required. One may then speculate about the two contributions
changing locally, i.e. the heavy quasiparticles related to
the bulk and the conduction band being related to the surface
states, both of which residing inside the hybridization gap.
Evidence for the former is found in specific heat
\cite{Gab01} and recent ARPES measurements \cite{Hla15}. The
latter have been detected in the de Haas-van Alphen (dHvA)
experiments, which report two dimensional Fermi surfaces
with light effective mass of the quasiparticles \cite{Li14}.

\section{Kondo effect}
As mentioned above, SmB$_6$ is an intermediate-valence
compound with a Sm valence of $\nu \approx 2.6$ \cite{Miz09}.
For such a value of $\nu$ it is questionable to what extent
the standard Kondo picture can be applied \cite{Var94}.

\begin{figure}[t]
\begin{center}
\includegraphics[width=13.4cm,clip]{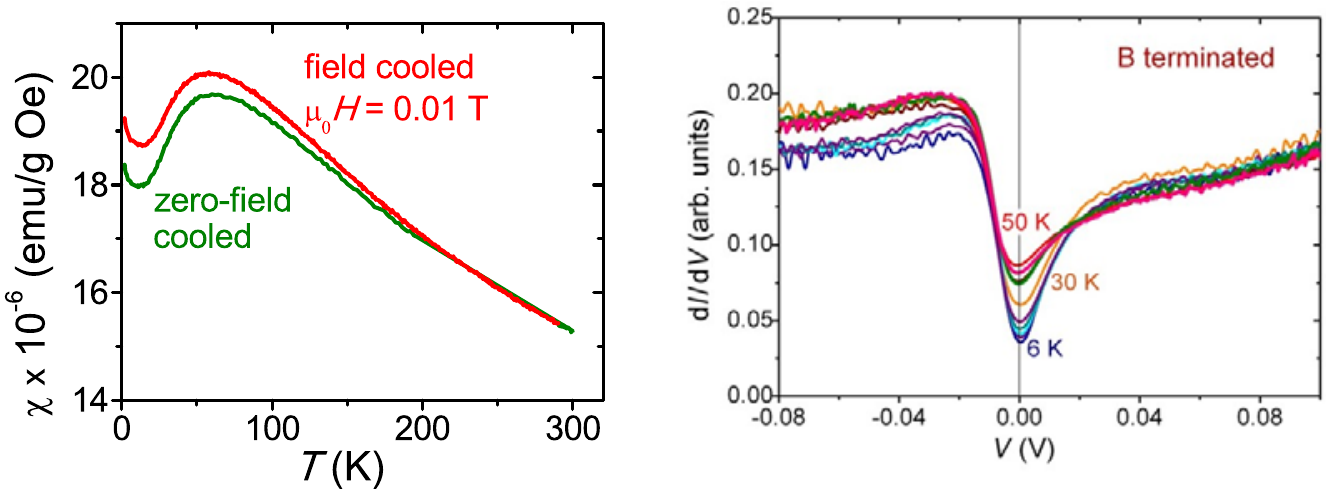}
\end{center}
\caption{Left: Temperature dependence of the susceptibility
$\chi$. Right: Tunneling spectra on a B-terminated surface
of SmB$_6$ at temperatures 6 K $\leq T \leq$ 50 K clearly
showing Fano-type line-shapes and a closing-up of the
partial gap at zero-bias with increasing $T$.}
\label{Kondo} \end{figure}

In figure \ref{Kondo} left, the susceptibility is recapped 
for one of our samples. In line with earlier measurements 
\cite{Zha13,Ris00,Gab02} it exhibits Curie-Weiss-like and
van Vleck contributions (due to its intermediate valence) 
for $T \gtrsim 100$ K, and non-magnetic behavior with an 
enlarged DOS at $E_{\rm F}$ at lowest temperatures.

Whether or not such a behavior is related to an underlying
Kondo effect has been discussed controversially, see e.g.
\cite{Aep92,Coo95,Col97,Ris00,Kas96}. More recent support
for the Kondo picture was provided e.g. by the analysis of
point contact measurements \cite{Zha13}. The closing-up of
the gap as observed by temperature dependent STS measurements
on small, non-reconstructed surface areas was also consistent
with expectations for the Kondo effect \cite{Roe14}. In
figure \ref{Kondo} right, $g(V)$-spectra obtained on a large,
non-reconstructed B-terminated area in the range 6 K $\leq T
\leq$ 50 K is presented. In contrast to \cite{Roe14}, the
spectra here are clearly of Fano-type. Yet, the closing of
the (partial) gap with increasing temperature follows the
same trend and hence, is again consistent with a logarithmic
behavior expected for the Kondo effect \cite{Cos00}. Note
that very similar Fano line-shapes were obtained on a ($2\!
\times \! 1$) reconstructed surface in the temperature range
8 K $\leq T \leq$ 30 K and the respective spectra were
interpreted in terms of hybridization resulting from the
underlying Kondo lattice \cite{Yee13}.

\section*{Acknowledgements}
The authors gratefully acknowledge experimental help by
Tae-Hwan Jang and insightful discussions with Yuri Grin. Work
at UC Irvine is supported by the Defense Advanced Research
Agency (DARPA) under agreement number FA8650-13-1-7374.

\end{document}